\documentclass[a4paper,nobibnotes,nofootinbib]{revtex4}

\usepackage{amsmath,amssymb}
\usepackage{epsfig}

\newcommand{\be}{\begin{equation}}
\newcommand{\ee}{\end{equation}}
\newcommand{\bL}{\begin{Large}}
\newcommand{\eL}{\end{Large}}
\newcommand{\ba}{\begin{eqnarray}}
\newcommand{\ea}{\end{eqnarray}}
\newcommand{\bc}{\begin{center}}
\newcommand{\ec}{\end{center}}
\newcommand{\bfig}{\begin{figure}}
\newcommand{\efig}{\end{figure}}

\begin{document}
\title{Hard diffraction at the LHC and the Tevatron using double pomeron exchange}

\author{C. Royon}\email{royon@hep.saclay.cea.fr}
\affiliation{Service de physique des particules, CEA/Saclay,
  91191 Gif-sur-Yvette cedex, France, and Fermilab, Batavia, USA}

\begin{abstract}
We use a Monte Carlo implementation of recently developped 
models of inclusive and exclusive diffractive $W$, top, Higgs and stop productions
to assess the sensitivity of the LHC experiments. We also discuss how the
Tevatron experiments could test the models and measure the gluon density in 
the pomeron, which is needed to make precise predictions at the LHC. 
\end{abstract}

\maketitle

Hard diffraction at the LHC has brought much interest recently related
to diffractive Higgs and SUSY production \cite{review}. It is thus important
that the different models available can be tested at the Tevatron before the
start of the LHC. In this contribution, we will consider only one model based on
the Bialas-Landshoff approach \cite{us}, and more details about other models and their
implications can be found in \cite{review} and the references therein.

\section{Theoretical framework}

We distinguish in the following the so called {\it inclusive} and 
{\it exclusive} models for
diffraction. We call exclusive models the models where almost the full energy available
in the center of mass is used to produce the heavy object (dijets, Higgs,
diphoton, $W$...). In other words, we get in the final state the diffractive
protons (which can be detected in roman pot detectors) and the heavy state
which decays in the main detector. The inclusive diffraction corresponds to
events where only part of the available energy is used to produce the heavy
object diffractively. For this model, we assume the pomeron is made of quarks
and gluons (we take the gluon and quark densities from the HERA measurements 
in shape and the normalisation from Tevatron data), and a quark or a gluon from
the pomeron is used to produce the heavy state. Thus the exclusive model appears
to be the limit where the gluon in the pomeron is a $\delta$ distribution in this
framework or in other words, there is no pomeron remnants in exclusive events.
We will see in the following that this distinction is quite relevant for
experimental applications.

\subsection{Exclusive model}
Let us first introduce the model \cite{bialas} we shall use for describing 
exclusive production. In \cite{bialas}, the diffractive mechanism is based on two-gluon 
exchange between the 
two incoming protons. The soft pomeron is seen as a  pair of gluons 
non-perturbatively coupled  to the proton. One of the gluons is then coupled 
perturbatively to the hard process while the other one plays the r\^ole of a 
soft screening of colour, allowing for diffraction to occur. We will give here the
formulae for either the SUSY Higgs boson, or 
the $\tilde t \bar{\tilde t}$ pairs production and other formulae for standard
model Higgs bosons, $t \bar{t}$, diphoton or dijet production can be found
in \cite{us}.
The corresponding cross-sections for Higgs bosons and $\tilde t \bar{\tilde t}$ 
production read:

\begin{eqnarray}
 d\sigma_{h}^{exc}(s) &=& C_{h}\left(\frac{s}{M_{h}^{2}}\right)^{2\epsilon} 
\delta\left(\xi_{1}\xi_{2}-\frac{M_{h}^{2}}{s}\right)
\prod_{i=1,2} \left\{ d^{2}v_{i} \frac{d\xi_{i}}{1-\xi_{i}} \right.
\left. \xi_{i}^{2\alpha'v_{i}^{2}} \exp(-2\lambda_{h} v_{i}^{2})\right\} 
\sigma (g g \rightarrow h)
 \nonumber \\
d\sigma_{\tilde t \tilde{\bar{t}}}^{exc}(s) &=& C_{\tilde t \tilde {\bar{t}}} 
\left(\frac{s}{M_{\tilde t \tilde{\bar{t}}^{2}}}\right)^{2\epsilon}
\delta\left( \sum_{i=1,2} (v_{i} + k_{i}) \right)
\prod_{i=1,2} \left\{ d^{2}v_{i} d^{2}k_{i} d\xi_{i} \right.
d\eta_{i}\  \xi_{i}^{2\alpha'v_{i}^{2}}\! 
\left. \exp(-2\lambda_{\tilde t\tilde{\bar{t}}} v_{i}^{2})\right\} {\sigma} 
(gg\to {\tilde t \tilde{\bar{t}}}\ ) \nonumber
\label{exclusif}
\end{eqnarray}
where, in both equations,  the variables $v_{i}$ and $\xi_{i}$ denote respectively 
the transverse
momenta and fractional momentum losses of the outgoing protons. In the second 
equation, 
$k_{i}$ and $\eta_{i}$ are respectively the squark  transverse 
momenta and  rapidities. $\sigma (g g \rightarrow H),  {\sigma} (gg\to {\tilde t 
\tilde{\bar{t}}}\ )$ are the hard  production cross-sections which are given 
later on. The model normalisation constants  $C_{h}, C_{\tilde t \tilde 
{\bar{t}}}$ are fixed from the fit to dijet diffractive production \cite{us},
while the ratio is fixed theoretically \cite{us}.

In the model,  the soft pomeron 
trajectory is
taken 
from the standard 
Donnachie-Landshoff   parametrisation \cite{donnachie},
 namely $\alpha(t) = 1 + 
\epsilon + \alpha't$, with
$\epsilon \approx 0.08$ and $\alpha' \approx 0.25 
\mathrm{GeV^{-2}}$. 
$\lambda_{h}, \lambda_{\tilde t \tilde{\bar{t}}}$ are  
kept as in  the original paper \cite{bialas} for the SM Higgs and $q 
\bar q$ pairs.  Note that, in this model, the 
strong (non perturbative) 
coupling constant is fixed to a reference value 
$G^2/4\pi,$ which will be taken 
from the fit to the observed centrally produced diffractive dijets.

In order to select exclusive diffractive states,  it is required to take into 
account the corrections from soft hadronic scattering. Indeed, the soft 
scattering  between incident particles tends to mask the genuine
hard diffractive interactions at 
hadronic colliders. The formulation of this 
correction \cite{alexander} to the
scattering amplitudes consists in considering a gap 
survival  probability.
The correction factor is commonly evaluated to be of order $0.03$ for the QCD 
exclusive diffractive processes at the LHC.

More details about the theoretical model and its phenomenological
applications can be found in Refs. \cite{ourpap} and \cite{us}. In the following,
we use the Bialas Landshoff model for exclusive Higgs production recently implemented in
a Monte-Carlo generator \cite{ourpap}. 

\subsection{Inclusive model}
Let us now discuss the inclusive models. We first notice that both models are
related, since they are both based on the Bialas Landshoff formalism.
The main difference, as we already mentionned, is that the exclusive model is a
limit of the inclusive model where the full energy available is used in the
interaction. The inclusive models implies the knowledge of the gluon and quark
densities in the pomeron. Whereas exclusive events are still to be observed,
inclusive diffraction has been studied already in detail at UA8 and then at HERA
and Tevatron. 

The inclusive mechanism is based on the 
idea that a Pomeron is a composite system, made itself from 
quarks and 
gluons. In our model, we thus apply the concept of Pomeron structure functions 
to 
compute the inclusive diffractive Higgs boson cross-section. The H1 
measurement of the diffractive structure function \cite{H1pom} and the 
corresponding quark and gluon 
densities are used for this purpose. This implies the existence of
Pomeron remnants and QCD radiation, as is the case for the proton. This 
assumption comes
from {\it QCD factorisation} of hard processes. 
However, and this is also an important issue, we do not assume {\it Regge 
factorisation} at the proton vertices, {\it i.e.} we do not use the H1 Pomeron 
flux factors in 
the proton or antiproton.

Regge factorisation is known to be violated between HERA and the
Tevatron. Moreover, we want to use the same physical idea as in the 
exclusive model \cite{bialas}, namely that a non perturbative gluon exchange 
describes the 
soft interaction between the incident particles. In 
practice, the 
Regge factorisation breaking appears in three ways in our model:

{\bf i)} We keep as in the original model of Ref \cite{bialas} the soft Pomeron
trajectory with an intercept value of 1.08.

{\bf ii)}  We normalize our
predictions to the CDF Run I measurements, allowing for factorisation breaking 
of the Pomeron flux factors in the normalisation between the HERA and hadron 
colliders \footnote{Indeed, recent results from a QCD fit to the diffractive 
structure 
function in H1 \cite{paplaurent} show that the discrepancy between the gluonic 
content of
the Pomeron at HERA and Tevatron appears mainly in
normalisation.}.

{\bf iii)} The color factor derives from the non-factorizable 
character of the model, since it stems from the gluon exchange between the 
incident hadrons. We will see later the difference between this and the 
factorizable
case.

The formulae for the inclusive production processes considered here follow. We 
have, 
for dijet production\footnote{We call ``dijets'' the produced quark and gluon 
pairs.}, 
considering only the dominant gluon-initiated hard processes:

\begin{eqnarray}
d\sigma_{JJ}^{incl} = C_{JJ}\left(\frac {x^g_1x^g_2 s }{M_{J
J}^2}\right)^{2\epsilon}\!\! \! \! \delta ^{(2)}\! 
\left( \sum _{i=1,2}
v_i\!+\!k_i\right) \prod _{i=1,2} \!\!\left\{{d\xi_i}  {d\eta_i}
d^2v_i d^2k_i {\xi _i}^{2\alpha' v_i^2}\!
\exp \left(-2 v_i^2\lambda_{JJ}\right)\right\} \times\nonumber \\
\times\left\{{\sigma_{JJ}} G_P(x^g_1,\mu) G_P(x^g_2,\mu) \right\};
\label{dinclujj}
\end{eqnarray}

\noindent and for Higgs boson production:

\begin{eqnarray}
d\sigma_H^{incl} = C_{H}\left(\frac {x^g_1x^g_2 s
}{M_{H}^2}\right)^{2\epsilon} \!\!\delta \left(\xi _1 \xi
_2\!-\!\frac{M_{H}^2}{x^g_1x^g_2 s} \right) \!\!\prod _{i=1,2}
\left\{G_P(x^g_i,\mu)\ dx^g_i  d^2v_i\ \frac {d\xi _i}{1\!-\!\xi 
_i}\
{\xi _i}^{2\alpha' v_i^2}\ \exp \left(-2 
v_i^2\lambda_H\right)\right\};
\label{dincluH}
\end{eqnarray}

\noindent In the above, the $G_P$ (resp. $Q_P$) are the Pomeron gluon (resp. 
quark) 
densities, and $x^{g}_{i}$ (resp. $x^{q}_{i}$) are the Pomeron's momentum 
fractions carried by 
the gluons (resp. quarks) involved in the hard process. We 
use as 
parametrizations of the Pomeron structure functions the fits to the diffractive 
HERA data 
performed in \cite{ba00, paplaurent}. Additional formulae concerning for instance
inclusive diffractive production of dileptons or diphotons are given in
\cite{us}.

Both the inclusive and exclusive productions have been implemented in a
generator called DPEMC, which has been interfaced with a fast simulation of the
D\O\ , CDF, ATLAS and CMS detectors.

\section{Experimental context}

In this section, we discuss mainly the parameters which we use to simulate the
detectors at the LHC. The simulation will be valid for both CMS and ATLAS
detectors.
The analysis is based on a fast simulation of the CMS detector at the LHC.
The calorimetric coverage of the CMS experiment ranges up to a pseudorapidity 
of $|\eta|\sim 5$. 
The region devoted
to precision measurements lies within $|\eta|\leq 3$, with a typical 
resolution on jet energy measurement of $\sim\!50\%
/\sqrt{E}$,
where $E$ is in GeV, and a granularity in pseudorapidity and azimuth of 
$\Delta\eta\times\Delta\Phi \sim 0.1\times 0.1$. 

In addition to the central CMS detector, the existence of roman pot detectors
allowing to tag diffractively produced protons,
located on both $p$ sides, is assumed \cite{helsinki}. The $\xi$ acceptance and 
resolution have been derived for each device using a complete simulation
of the LHC beam parameters. The combined $\xi$ acceptance is $\sim 100\%
$ 
for $\xi$ ranging from $0.002$ to $0.1$, where
$\xi$ is 
the proton fractional momentum loss. The acceptance limit of the device 
closest to the interaction point
is $\xi > \xi_{min}=$0.02. 

In exclusive double Pomeron exchange, the mass of the central 
heavy object is given by $M^2 = \xi_1\xi_2 s$, where $\xi_1$ and $\xi_2$ are
the proton fractional momentum losses measured in the roman pot detectors.
At this level, we already see the advantages of the exclusive events. Since,
there is no energy loss due to additional radiation or pomeron remnants, we can
reconstruct the total diffractive mass, which means the mass of the
diffractively produced object (the Higgs, dijets, $t \bar{t}$, 
$t \tilde{\bar{t}}$, events, $W$ pairs...), very precisely using the kinematical
measurements from the roman pot detectors. The mass resolution is thus coming
directly from the $\xi$ resolution which is expected to be of the order of 1\%.
For inclusive events, the mass resolution will not be so good since part of the
energy is lost in radiation, which means that we measure the mass of the heavy
object produced diffractively and the pomeron remnants together very precisely.
To get a good mass resolution using inclusive events requires a good
measurement
of the pomeron remnants and soft radiation and being able to veto on it.

\section{Existence of exclusive events}
While inclusive diffraction has already been observed at many colliders,
the question arises whether exclusive events exist or not since they have never been
observed so far. This is definitely an area where the Tevatron experiments can
help to test the models and show evidence for the existence of exclusive events
if any. It is crucial to be able to test the different models before the 
start of the LHC.
The D\O\ and CDF experiments 
at the Tevatron (and the LHC experiments) are ideal places to look for
exclusive events in dijet or $\chi_C$ channels for instance
where exclusive events are expected to occur at high dijet mass
fraction.
So far, no evidence of the existence of exclusive events has been found.
The best way to show evidence of the existence of exclusive events would be the
measurement of the ratio of the diphoton to the dilepton cross sections
as a function of the diphoton/dilepton mass ratio (the diphoton-dilepton mass
ratio being defined as the diphoton-dilepton mass divided by the total
diffractive mass).
The reason is quite simple: it is possible to produce exclusively diphoton but
not dilepton directly since ($g g \rightarrow \gamma \gamma$) 
is possible but not  ($g g \rightarrow l^+ l^-$) directly at leading
order. The ratio of
the diphoton to the dilepton cross section should show a bump or a change of
slope towards high diphoton-dilepton masses if exclusive events exist.
Unfortunately, the production cross section of such events is small and it will
probably not be possible to perform this study before the start of the LHC.

Another easier way to show the existence of such events would be to study the
correlation between the gap size measured in both $p$ and $\bar{p}$ directions
and the value of $\log 1/\xi$ measured using roman pot detectors, which can be
performed in the D\O\ experiment. The gap size between the
pomeron remnant and the protons detected in roman pot detector 
is of the order of 
$log 1/\xi$ for usual diffractive events (the measurement giving a slightly
smaller value to be in the acceptance of the forward detectors) while
exclusive events show a much higher value for the rapidity gap since the gap
occurs between the jets (or the $\chi_C$) and the proton detected in roman
pot detectors (in other words, there is no pomeron remnant)
\footnote{To distinguish between pure exclusive and
quasi-exclusive events (defined as inclusive diffractive events where little
energy is taken away by the pomeron remnants, or in other words, events
where the mass of the heavy object produced diffractively is almost equal
to the total diffractive mass), other observables such as
the ratio of the cross sections of double diffractive
production of diphoton and dilepton, or the $b$-jets to all jets 
are needed \cite{us}.}. Another observable leading
to the same conclusion would be the correlation between $\xi$ computed
using roman pot detectors and using only the central detector.

Another way to access the existence of exclusive events would be via QCD
evolution. If one assumes that the DGLAP evolution equations work for parton
densities in the pomeron, it is natural to compare the predictions of
perturbative QCD with for instance dijet production in double pomeron exchange
as a function of the dijet mass fraction (defined as the ratio of the dijet mass
divided by the total diffractive mass) for different domains in diffractive
mass. It has been shown that the dependence of the exclusive production cross
section as a function of the dijet mass is much larger than the one of the
inclusive processes. In other words, if exclusive events exist, it is expected
that the evolution of the dijet cross section in double pomeron exchanges  as a
function of dijet mass fraction in bins of dijet masses will be incompatible
with standard QCD DGLAP evolution, and will require an additional contribution,
namely the exclusive ones \footnote{Let us note that one should also
distinguish this effect from higher order corrections, and also from higher
twist effects, which needs further studies.}. 
It will be quite interesting to perform such
an analysis at the Tevatron if statistics allows.

\section{Results on diffractive Higgs production}
Results are given in Fig. 1 for a Higgs mass of 120 GeV, 
in terms of the signal to background 
ratio S/B, as a function of the Higgs boson mass resolution.
Let us notice that the background is mainly due the exclusive $b \bar{b}$
production. However the tail of the inclusive $b \bar{b}$ production can also be
a relevant contribution and this is related to the high $\beta$ gluon
density which is badly known as present. It is thus quite important to constrain
these distributions using Tevatron data as suggested in a next section.

In order to obtain an S/B of 3 (resp. 1, 0.5), a mass resolution of about
0.3 GeV (resp. 1.2, 2.3 GeV) is needed. The forward detector design of 
\cite{helsinki} 
claims a resolution of about 2.-2.5 GeV, which leads to a S/B of about 
0.4-0.6. Improvements in this design
would increase the S/B ratio as indicated on the figure.
As usual, this number is enhanced by a large factor if one considers 
supersymmetric Higgs boson 
production with favorable Higgs or squark field mixing parameters.

The cross sections obtained after applying the survival probability of 3\% at
the LHC as well as the S/B ratios are given in Table \ref{sb} if one assumes a
resolution on the missing mass of about 1 GeV (which is the most optimistic
scenario). The acceptances of the roman pot detectors as well as the simulation
of the CMS detectors have been taken into account in these results. 

Let us also notice that the missing mass method will allow to perform a $W$ 
mass measurement using exclusive (or quasi-exclusive) $WW$ 
events in double Pomeron exchanges, and QED processes
\cite{ushiggs}. The advantage of the
QED processes is that their cross section is perfectly known and that this
measurement only depends on the mass resolution and the roman pot acceptance.
In the same way, it is possible to measure the mass of the top quark in
$t \bar{t}$ events in double Pomeron exchanges \cite{ushiggs} 
as we will see in the following.

\begin{table}
\begin{center}
\begin{tabular}{|c||c|c|c|c|c|} \hline
$M_{Higgs}$& cross & signal & backg. & S/B & $\sigma$  \\
 & section &  & & & \\
\hline\hline
120 & 3.9 & 27.1 & 28.5 & 0.95 & 5.1  \\
130 & 3.1 & 20.6 & 18.8 & 1.10 & 4.8  \\
140 & 2.0 & 12.6 & 11.7 & 1.08 & 3.7  \\ 
\hline
\end{tabular}
\caption{Exclusive Higgs production cross section for different Higgs masses,
number of signal and background events for 100 fb$^{-1}$, ratio, and number of
standard deviations ($\sigma$).}
\label{sb}
\end{center}
\end{table}

The diffractive SUSY Higgs boson production cross section is noticeably enhanced 
at high values of $\tan \beta$ and since we look for Higgs decaying into $b
\bar{b}$, it is possible to benefit directly from the enhancement of the cross
section contrary to the non diffractive case. A signal-over-background up to a
factor 50 can be reached for 100 fb$^{-1}$ for $\tan \beta \sim 50$
\cite{lavignac}. We give in Figure 2 the
signal-over-background ratio for different values of $\tan \beta$ for a Higgs
boson mass of 120 GeV.


\begin{figure}[htb]
\begin{center}
\includegraphics[width=11.5cm,clip=true]{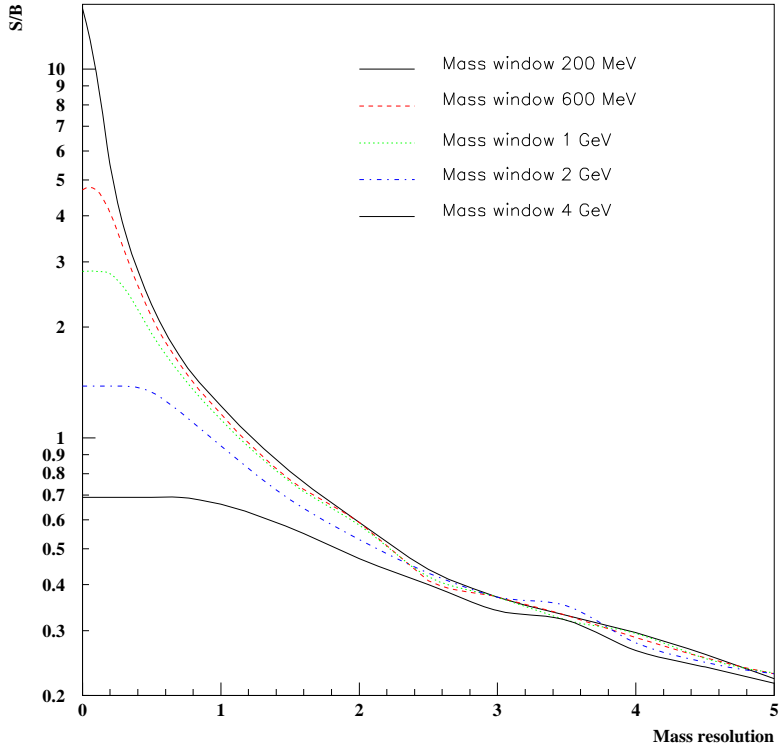}
\caption{Standard Model Higgs boson signal to background ratio as a function 
  of the resolution on
        the missing mass, in GeV. This figure assumes a Higgs
        boson mass of 120 GeV.}
\end{center}
\end{figure}

\begin{figure}[htb]
\begin{center}
\includegraphics[width=11.5cm,clip=true]{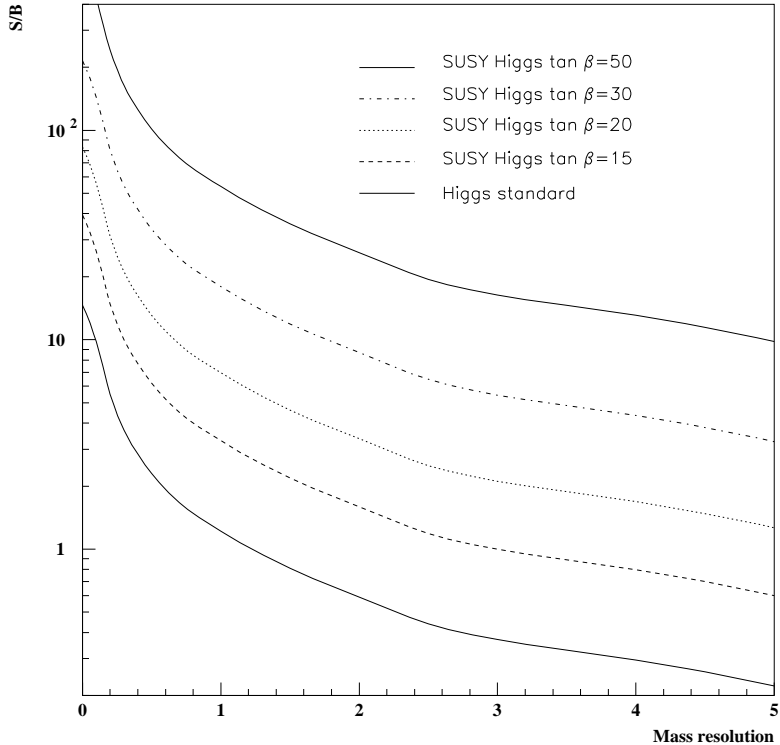}
\caption{SUSY Higgs boson signal to background ratio as a function 
  of the resolution on
        the missing mass, in GeV. This figure assumes a Higgs
        boson mass of 120 GeV.}
\end{center}
\end{figure}

\section{Threshold scan method: $W$, top and stop mass measurements}
We propose a new method to measure heavy particle properties via double 
photon and double pomeron exchange (DPE), at the LHC \cite{ushiggs}. In this category of events, the heavy objects 
are produced in pairs, whereas the beam particles
often leave the interaction region intact, and can be measured using very forward detectors.

Pair production of $WW$ bosons and top quarks in QED and  double pomeron exchange are described in detail in this section. 
$WW$ pairs are produced in photon-mediated processes, which are exactly calculable in QED. There is 
basically no uncertainty concerning the possibility of measuring these processes
at the LHC. On the contrary, $t \bar{t}$ events, produced in 
exclusive double pomeron exchange, suffer from theoretical uncertainties since 
exclusive diffractive production is still to be observed at the Tevatron, 
and other models lead to different cross sections, and thus to a different
potential for the top quark mass measurement. However, since the exclusive 
kinematics are simple, the model dependence will be essentially reflected by 
a factor in the effective luminosity for such events.

\subsection{Explanation of the methods}
We study two different methods to reconstruct the mass of heavy objects
double diffractively produced at the LHC. The method is
based on a fit to the turn-on point of the missing mass distribution at 
threshold. 

One proposed method (the ``histogram'' method) corresponds to the comparison of 
the mass distribution in data with some reference distributions following
 a Monte Carlo simulation of the detector with different input masses
corresponding to the data luminosity. As an example, we can produce 
a data sample for 100 fb$^{-1}$ with a top mass of 174 GeV, and a few 
MC samples corresponding to top masses between 150 and 200 GeV by steps of. 
For each Monte Carlo sample, a $\chi^2$ value corresponding to the 
population difference in each bin between data and MC is computed. The mass point 
where
the $\chi^2$ is minimum corresponds to the mass of the produced object in data.
This method has the advantage of being easy but requires a good
simulation of the detector.

The other proposed method (the ``turn-on fit'' method) is less sensitive to the MC 
simulation of the
detectors. As mentioned earlier, the threshold scan is directly sensitive to
the mass of the diffractively produced object (in the $WW$W case for instance, it
is sensitive to twice the $WW$ mass). The idea is thus to fit the turn-on
point of the missing mass distribution which leads directly to the mass 
of the produced object, the $WW$ boson. Due to its robustness,
this method is considered as the ``default" one in the following.

\subsection{Results}

To illustrate the principle of these methods and their achievements,
we  apply them to the 
$WW$ boson and the top quark mass measurements in the
following, and obtain the reaches at the LHC. They can be applied to other 
threshold scans as well.
The precision of the $WW$ mass measurement (0.3 GeV for 300 fb$^{-1}$) is not competitive with other 
methods, but provides a very precise calibration 
of the roman pot detectors. The precision of
the top mass measurement is however competitive, with an expected precision 
better than 1 GeV at high luminosity. The resolution on the top mass is given
in Fig. 3 as a function of luminosity for different resolutions of the roman
pot detectors.

The other application is to use the so-called ``threshold-scan method"
to measure the stop mass in {\it exclusive} events. The idea is straightforward: 
one
measures the turn-on point in the missing mass distribution at about twice
the stop mass. After taking into account the stop width, we obtain a resolution
on the stop mass of 0.4, 0.7 and 4.3 GeV for a stop mass of 174.3, 210 and 393
GeV for a luminosity (divided by the signal efficiency) of 100 fb$^{-1}$. We
notice that one can expect to reach typical mass resolutions which can be obtained at a linear
collider. The process is thus similar to  those at linear colliders (all final 
states
are detected) without the initial state radiation problem. 

The caveat is  of course that production via diffractive 
{\it exclusive} processes is model dependent, and definitely needs
the Tevatron data to test the models. It will allow to determine more precisely 
the production cross section by testing and measuring at the Tevatron the jet 
and photon production for high masses and high dijet or diphoton mass fraction.

\section{How to constrain the high $\beta$ gluon using Tevatron and LHC data?}

In this section, we would like to discuss how we can measure the gluon density
in the pomeron, especially at high $\beta$ since the gluon in this kinematical 
domain shows large uncertainties and this is where the exclusive contributions
should show up if they exist. To take into account, the high-$\beta$
uncertainties of the gluon distribution, we chose to multiply the gluon density
in the pomeron measured at HERA by a factor $(1-\beta)^{\nu}$  where $\nu$
varies between -1.0 and 1.0. If $\nu$ is negative, we enhance the gluon density
at high $\beta$ by definition, especially at low $Q^2$.

The measurement of the dijet mass fraction at the Tevatron for two jets with a
$p_T$ greater than 25 GeV for instance in double pomeron exchange is indeed
sensitive to these variations in the gluon distribution. The dijet mass fraction
is given in Fig 3
and 4 which shows how the Tevatron data can
effectively constrain the gluon density in the pomeron \cite{paplaurent}.
Another possibility to measure precisely the gluon distribution in the pomeron
at high $\beta$ would be at the LHC the measurement of the $t \bar{t}$ cross
section in double pomeron exchange in inclusive events
\cite{papttbarincl}. By requiring the
production of high mass objects, it is possible to assess directly the tails of
the gluon distribution. In Fig.5, we give the total mass
reconstructed in roman pot detectors for double tagged events in double pomeron
exchanges and the sensitivity to the gluon in the pomeron.

\begin{figure}[htb]
\begin{center}
\includegraphics[width=11.5cm,clip=true]{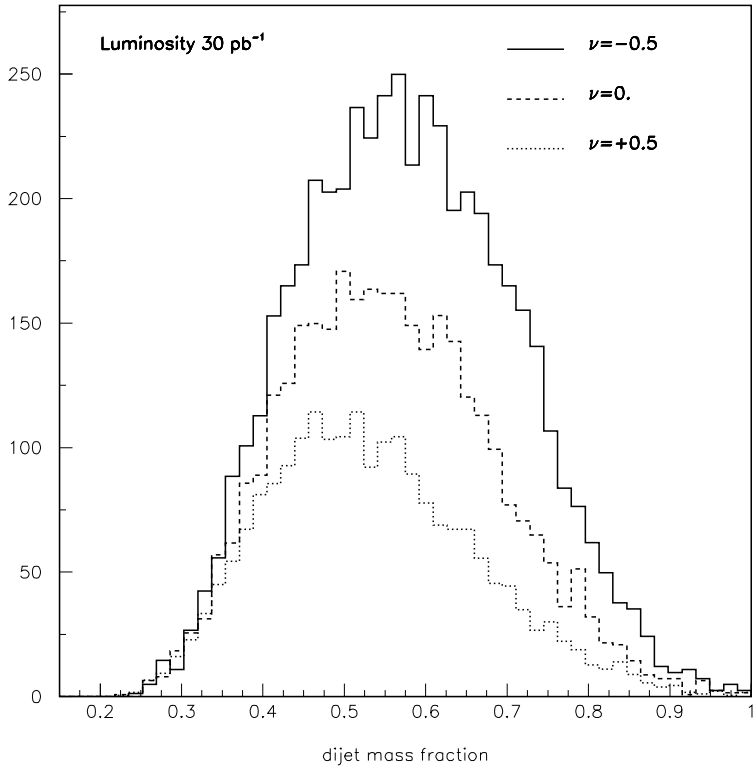}
\caption{Dijet mass fraction at the Tevatron at generator level when the gluon
density measured in the H1 experiment is used 
\cite{ba00, paplaurent} and multiplied by $(1 -
\beta)^{\nu}$. We notice the sensitivity of this measurement on the gluon 
density.}
\end{center}
\label{dijetmassfraction}
\end{figure}

\begin{figure}[htb]
\begin{center}
\includegraphics[width=11.5cm,clip=true]{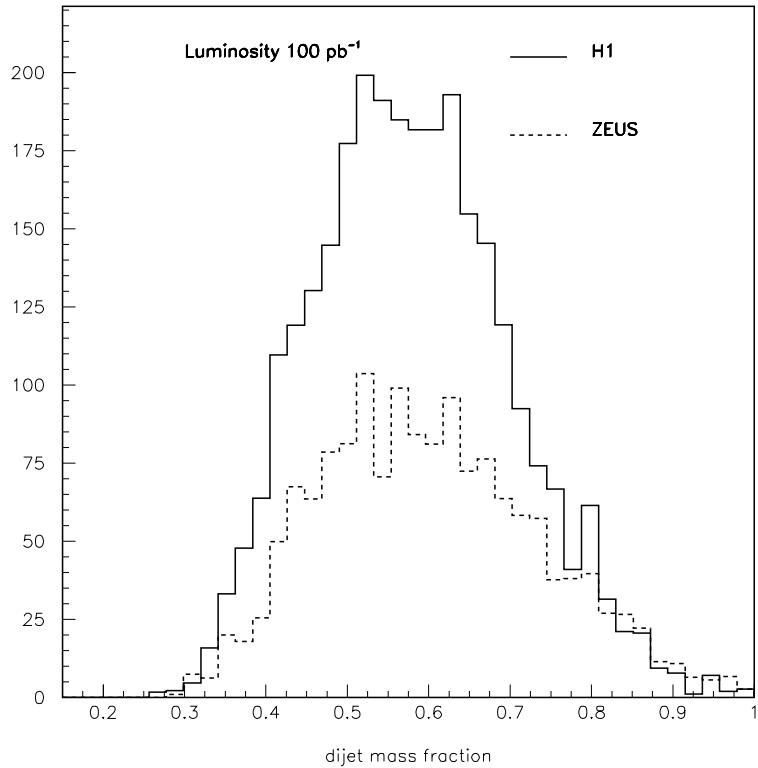}
\caption{Dijet mass fraction at the Tevatron at generator level when the gluon
density measured in the H1 or the ZEUS experiment is used \cite{ba00, paplaurent}
and one tag
is asked in the roman pot acceptance of the D\O\ or the CDF collaboration in the
$\bar{p}$ direction. We notice the sensitivity of this measurement on the gluon 
density.}
\end{center}
\label{dijetmassfractionb}
\end{figure}

\begin{figure}[htb]
\begin{center}
\includegraphics[width=11.5cm,clip=true]{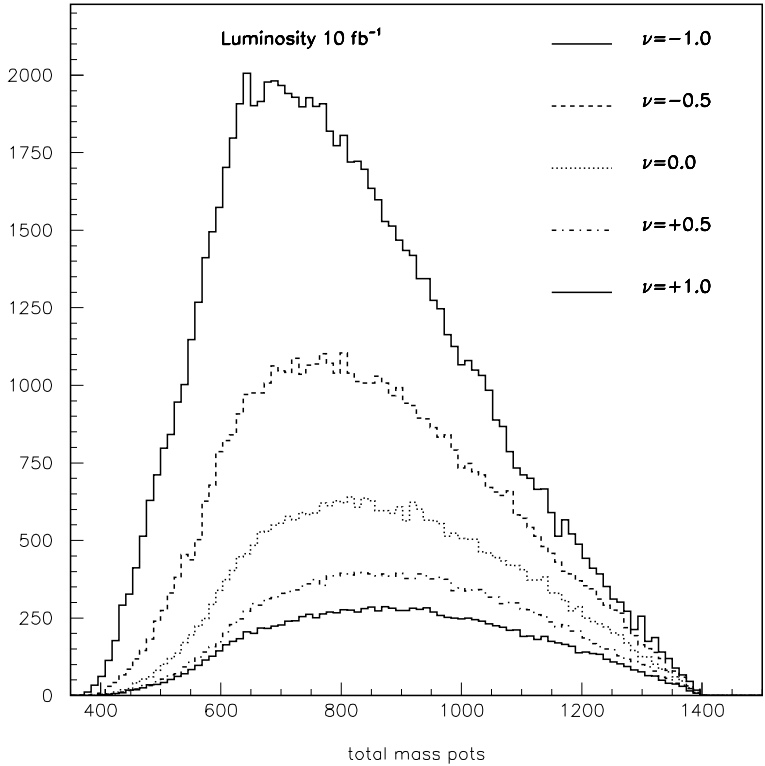}
\caption{Total diffractive mass reconstructed 
for $t \bar{t}$ inclusive events in double pomeron exchanges
using roman detectors at the LHC. We use the gluon density in the pomeron
measured in the H1 experiment \cite{ba00, paplaurent} and we multiply it by $(1 -
\beta)^{\nu}$ to show the sensitivity on the gluon density at high $\beta$.}
\end{center}
\label{ttbarincl}
\end{figure}

\section{New possible measurement of survival probabilities in the D\O\
experiment}

We propose a new measurement to be performed at the Tevatron,
in the D\O\ experiment \cite{alexander}, which can be
decisive to distinguish between Pomeron-based and soft colour interaction
models of hard diffractive scattering. This measurement allows
to test directly the survival probability parameters as well which is
fundamental to predict correctly the exclusive diffractive Higgs production at
the LHC.
The discriminative potential of our proposal takes its origin  in the 
factorization breaking properties which were already observed at the 
Tevatron. The explanation given to this factorization 
breaking in Pomeron-based models is the occurrence of large corrections 
from the  survival 
probabilities, which is the probability to keep a diffractive event
signed either by tagging the proton in the final state or by requiring the
existence of a rapidity gap in the event. By contrast with Pomeron models, 
the soft color interaction models are by nature, non factorizable. The initial 
hard interaction is the generic standard dijet production, accompanied by 
the full radiative partons. Then, a phenomenological soft color interaction is assumed 
to modify the overall color content, allowing for a probability of color singlet 
exchange and thus diffraction.

The forward detector apparatus 
in the D\O\ experiment at the Tevatron, Fermilab, has the 
potential to discriminate between the predictions of the two approaches in  
hard ``double'' diffractive production, e.g. of  centrally produced dijets, 
by looking 
to the azimuthal 
distributions of the outgoing proton and antiproton with respect to the 
beam direction. This measurement relies on tagging  
both outgoing particles in roman pot detectors
installed by  the D0 experiment. 

The FPD consists of 
eight {\it quadrupole} spectrometers, four being located on the
outgoing proton side, the other four on the antiproton side. On each side,  
the quadrupole spectrometers are
placed both in the inner (Q-IN), and outer (Q-OUT) sides of the accelerator ring,
as well as in the upper (Q-UP) and lower (Q-DOWN) directions.
They provide almost full coverage in azimuthal angle $\Phi$.
The {\it dipole} spectrometer, marked as D-IN, is placed in the inner side of 
the ring, 
in the direction of outgoing antiprotons.

Each spectrometer allows to reconstruct the
trajectories of outgoing protons and antiprotons near the beam pipe
and thus to measure their energies and scattering angles. Spectrometers provide
high precision measurement in $t = -p_{T}^2$ and $\xi = 1 - p^\prime /
p$ variables,
where $p^\prime$ and $p_T$ are the total and transverse momenta of the outgoing 
proton or antiproton, and $p$ is the beam energy.
The dipole detectors show a good acceptance
down to $t=0$  for $\xi > 3. 10^{-2}$ and the quadrupole detectors
are sensitive on outgoing particles down to $|t|= 0.6$ GeV$^2$ for
$\xi < 3. 10^{-2}$, which allows to get a good acceptance for
high mass objects produced diffractively in the D\O\ main detector. 
For our analysis, we use a full simulation of 
the FPD acceptance in $\xi$ and $t$ \cite{fpdaccep}.

We suggest to count the number of events
with tagged $p$ and $\bar{p}$ for different combinations of FPD spectrometers.
For this purpose, we define the following
configurations for dipole-quadrupole tags
(see Fig. 2): same side (corresponding to D-IN on $\bar{p}$ side
and Q-IN on $p$ side and thus to $\Delta \Phi < 45$ degrees), 
opposite side (corresponding to D-IN on $\bar{p}$ side
and Q-OUT on $p$ side, and thus to $\Delta \Phi > 135$ degrees),
and middle side (corresponding to D-IN on $\bar{p}$ side
and Q-UP or Q-DOWN on $p$ side and thus to $45 < \Delta \Phi < 135$ degrees).
We define the same kinds of configurations for quadrupole-quadrupole tags
(for instance, the same side configuration corresponds to sum of the four
possibilities: both protons and antiprotons tagged in Q-UP, Q-DOWN,
Q-IN or Q-OUT).

In Table 2, we give the ratios $middle/(2 \times same)$ and $opposite/same$ 
(note that we divide $middle$ by 2 to get the same domain size in
$\Phi$) for the different models. In order to obtain these predictions, we used
the full acceptance in $t$ and $\xi$ of the FPD detector \cite{fpdaccep}.
Moreover we computed the ratios for two  different tagging configurations
namely for $\bar{p}$ tagged
in dipole detectors, and $p$ in quadrupoles, or for both $p$ and $\bar{p}$ 
tagged in quadrupole detectors.

In Table 2, we notice that the $\Phi$ dependence of the
event rate ratio for the SCI \cite{sci} model 
is weak, whereas for the POMWIG \cite{sci} models the result show important differences 
specially when both $p$ and $\bar{p}$ are tagged in quadrupole
detectors. This measurement can be performed even at low luminosity (1 week of
data) 
if two jets with a transverse momentum greater than 5 GeV are required.

With more luminosity, we also propose to measure directly the $\Delta \Phi$ dependence
between the outgoing protons and antiprotons using the good coverage of the
quadrupole detectors in $\Phi$ which will allow to perform a more
precise test of the models.

\begin{table}
\begin{center}
\begin{tabular}{|c|c||c|c|} \hline
 Config. & model & midd./ & opp./ \\ 
         &       &  same  & same \\
\hline\hline
Quad.  & SCI & 1.3 & 1.1 \\
 $+$ Dipole              & Pom. 1 & 0.36 & 0.18 \\
               & Pom. 2 & 0.47 & 0.20 \\ \hline
Quad.  & SCI & 1.4 & 1.2 \\
$+$ Quad.               & Pom. 1 & 0.14 & 0.31 \\
               & Pom. 2 & 0.20 &  0.049     
\\
\hline
\end{tabular}
\end{center}
\caption{Predictions for a proposed measurement of diffractive 
cross section ratios in different regions of $\Delta \Phi$ at the Tevatron
(see text for the definition of middle, same and opposite).
The first (resp. second) measurement involves the quadrupole and dipole
detectors (resp. quadrupole detectors only) leading to asymmetric (resp.
symmetric) cuts on $t$.
We notice that the SCI models do not predict any significant dependence
on $\Delta \Phi$ whereas the Pomeron-based models show large variations. }
\end{table}

\begin{figure}[htb]
\includegraphics[width=12.5cm,clip=true]{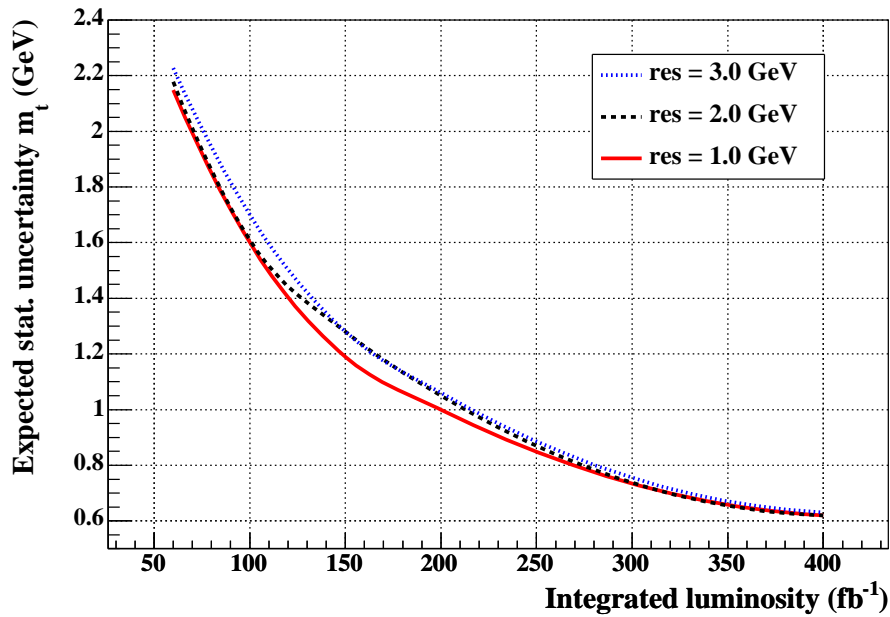}
\caption{Expected statistical precision of the top mass
    as a function of the integrated luminosity for various resolutions
    of the roman pot detectors (full line: resolution of 1 GeV, dashed line: 2
    GeV, dotted line: 3 GeV).}
\end{figure}

\section*{Acknowledgments}
There results come from a fruitful collaboration with M. Boonekamp, J. Cammin,
A. Kupco,
S. Lavignac, R. Peschanski and L. Schoeffel.

\end{document}